\documentstyle[12pt]{article}
\title{
Recursive Method for the Solution \\
of Systems of Linear Equations 
\thanks{
This paper vas published in: 
{\em Computational Mathematics} (A. Sydow Ed, Proceedings of the
15th IMACS World Congress, Vol. I, Berlin, August 1997 ),
Wissenschaft \& Technik Verlag, Berlin 1997, 475--480. 
No part of this materials may be reproduced, stored in retrieval system,
or transmitted, in any form without prior permission of the copyright owner.
}}

\author{
Gennadi.I.Malaschonok\\
Tambov State University, 392622 Tambov, Russia\\
e-mail: \ malaschonok@math-univ.tambov.su}
\date{}

\begin{document}

\maketitle

\abstract{
New solution method for the systems  of  linear  equations  in
commutative integral domains is proposed. Its complexity is
the same that the complexity of the matrix multiplication.}

\section{Introduction}

One of the first results in the theory of computational complexity
is the Strassen discovery of the new algorithm for matrix multiplication
\cite{str-1969}.
He changed the classical method with the complexity $O(n^3)$
for the new algorithm with the complexity $O(n^{\log_27})$. This
method may be used for a matrix in any {\it commutative ring}. He used
matrix multiplication for the computation of the inverse matrix,
of the determinant of a matrix and for the solution of the systems
of linear equations over an {\it arbitrary field} with the complexity
$O(n^{\log_27})$.

Many authors improved this result. There is known now  an algorithm
of matrix multiplication with the complexity $O(n^{2,37})$ (see
D.Coppersmith, S.Winograd \cite{cw-1987}).

We have another situation with the problems of the solution of systems
of linear equations and of the determinant computation in the
{\it commutative rings}. Dodgson \cite{dod-1866} proposed a method for the
determinant computation and the solution of systems of linear equations over
the ring of integer numbers with the complexity $O(n^3)$. During this century
this result was improved and generalized for arbitrary commutative
integral domain due to Bareis \cite{bar-1968} and the author (see
\cite{mal-1983} -- \cite{mal-1991}).
But the complexity is still $O(n^3)$.

There is proposed the new solution method for the systems of linear
equations in integral domains. Its complexity is
the same that the complexity of the matrix multiplication in integral domain.

Let
$$ \sum_{j=1}^{m-1} a_{ij} x_j = a_{im},~~i=1,2,\ldots,n $$
be the system of linear  equations with extended coefficients matrix
$$A=(a_{ij}), \; i=1, \ldots, n, \; j=1, \ldots, m.$$
whose coefficients are in integral domain ${\bf R}$:
$A~ \in ~{\bf R}^{n\times m}.$

The solution of  such  system  may  be  written  according  to
Cramer's rule
$$x_{j}={\delta^n_{jm}-\sum ^{m-1}_{p=n+1}x_p
\delta^{n}_{jp} \over \delta^n}, \; j=1, \ldots, n,$$
where $x_{p}$, $p=n+1, \ldots, m$,
are free variables and $\delta^{n}\neq 0$. $\delta ^{n}=|a_{ij}|$,
$i=1, \ldots, n$, $j=1, \ldots, n$,  -  denote the
corner minors of the matrix $A$ of order $n, \delta ^{n}_{ij}$  -
denote  the minors obtained by a substitution of the column $j$ of the matrix
$A$ instead of the column $i$ in the minors $\delta ^{n}$, $i=1, \ldots, n$,
$j=n+1, \ldots, m$. So we  need  to construct the algorithm of computation of
the minor $\delta ^{n}$ and the matrix
$G=\pmatrix{\delta ^{n}_{ij}\cr}$, $i=1, \ldots, n$, $j=n+1,n+2,\ldots, m$.

That means that we must make the reduction of the matrix $A$
to the diagonal form
$$A \rightarrow (\delta^n I_n, G).$$
$I_n$ denotes the unit matrix of order $n$.

\section{Recursive Algorithm}

For the extended coefficients matrix $\bf A$ we shall denote:
$$
{\bf A}_{ij}^k = \pmatrix{
a_{11}&a_{12}&\cdots&a_{1,k-1}&a_{1j}\cr
a_{21}&a_{22}&\cdots&a_{2,k-1}&a_{2j}\cr
\vdots&\vdots&\ddots&\vdots&\vdots\cr
a_{k-1,1}&a_{k-1,2}&\cdots&a_{k-1,k-1}&a_{k-1,j}\cr
a_{i1}&a_{i2}&\cdots&a_{i,k-1}&a_{ij}\cr
}$$
-- the matrix, formed by the surrounding of the submatrix
of an order $k-1$ in the upper left corner by row $i$ and column $j$,
$$a_{ij}^k= \det {\bf A}_{ij}^k,$$
$a_{ij}^1= a_{ij}$, $\delta^0=1$, $\delta^k =a_{kk}^k$, $\delta^k_{ij}$ --
the determinant of the matrix, that is received from the matrix
${\bf A}_{kk}^k$ after the substitution of the column $i$ by the column $j$.

We shall use the minors $\delta _{ij}^k$ and  $a _{ij}^k$ for the construction
of the matricez
$$
A_{k,c}^{r,l,(p)}=
\pmatrix{
a^p_{r+1,k+1}&a^p_{r+1,k+2}&\cdots&a^p_{r+1,c}\cr
a^p_{r+2,k+1}&a^p_{r+2,k+2}&\cdots&a^p_{r+2,c}\cr
\vdots&\vdots&\ddots&\vdots\cr
a^p_{l,k+1}&a^p_{l,k+2}&\cdots&a^p_{l,c}\cr
}$$
and
$$
G_{k,c}^{r,l,(p)}=
\pmatrix{
\delta^p_{r+1,k+1}&\delta^p_{r+1,k+2}&\cdots&\delta^p_{r+1,c}\cr
\delta^p_{r+2,k+1}&\delta^p_{r+2,k+2}&\cdots&\delta^p_{r+2,c}\cr
\vdots&\vdots&\ddots&\vdots\cr
\delta^p_{l,k+1}&\delta^p_{l,k+2}&\cdots&\delta^p_{l,c}\cr
}$$
$G_{k,c}^{r,l,(p)}, A_{k,c}^{r,l,(p)} \in {\bf R}^{(l-r)\times (c-k)}$,
$0 \leq k<n$, $k<c \leq n $, $0 \leq r<m$, $r<l \leq m $, $1 \leq p \leq n$.

We shall describe one recursive step, that makes the following reduction
of the matrix $\tilde A$ to the diagonal form
$$
\tilde A \rightarrow (\delta^{l} I_{l-k}, \hat G)
$$
where
$$
\tilde A= A_{k,c}^{k,l,(k+1)},\;\;\; \hat G= G_{l,c}^{k,l,(l)}
$$
$0 \leq k < c \leq m$, $k<l \leq n $, $l<c$.
Note that if $k=0$, $l=n$ and $c=m$ then we get the solution
of the system.

We can choose the arbitrary integer number $s$: $k< s <l$ and
write the matrix $\tilde A$ as the following:
$$
\tilde A=\pmatrix{A^1 \cr A^2 \cr}
$$
where $A^1=A_{k,c}^{k,s,(k+1)}$ - the upper part  of the matrix $\tilde A$
consists of the $s-k$ rows
and
$A^2= A_{k,c}^{s,l,(k+1)}$ - the  lower  part  of  the
matrix $\tilde A$.

\subsection{The first step}

As the next recurcive step we make the following reduction
of the matrix $A^1\in {\bf R}^{(s-k)\times (c-k)}$ to the diagonal form
$$
A^1 \rightarrow (\delta^{s} I_{s-k}, G_2^1),
$$
where $G_2^1= G_{s,c}^{k,s,(s)}$.

\subsection{The second step}

We write the matrix $A^2$ in the following way:
$$
A^2=(A_1^2,A_2^2)
$$
where $A_1^2= A_{k,s}^{s,l,(k+1)}$ consists of the first $s-k$ columns
and $A_2^2= A_{s,c}^{s,l,(k+1)}$ consists of the last $c-s$
columns of the matrix $A^2$.

The matrix $\hat A_2^2=A_{s,c}^{s,l,(s+1)}$
is obtained from the matrix identity (see the proof in the next section):
$$
\delta^k \cdot \hat A_2^2= \delta^{s} \cdot A_2^2-A_1^2 \cdot G_2^1.
$$
The minors $\delta^k$ must not equal zero.

\subsection{The third step}

As the next recurcive step we make the following reduction
of the matrix $\hat A^2_2 \in {\bf R}^{(l-s)\times (c-s)}$ to the
diagonal form
$$
\hat A^2_2 \rightarrow (\delta^{l} I_{l-s}, \hat G_{2''}^2),
$$
where $ \hat G_{2''}^2= G_{l,c}^{s,l,(l)}$.

\subsection{The fourst step}

We write the matrix $G_2^1$ in the following way:
$$
G_2^1=(G_{2'}^1, G_{2''}^1)
$$
where $G^1_{2'}= G_{s,l}^{k,s,(s)}$ consists of the first $l-s$ columns
and $G^1_{2''}= G_{l,c}^{k,s,(s)}$ consists of the last $c-l$
columns of the matrix $G_2^1$.

The matrix $\hat G_{2''}^1=
G_{l,c}^{k,s,(l)}$ is obtained from the matrix identity (see the proof
in the next section):
$$
\delta^{s} \cdot \hat G_{2''}^1= \delta^{l} \cdot G_{2''}^1-G_{2'}^1
\cdot \hat G_{2''}^2.
$$
The minors $\delta^{s}$ must not equal zero.

So we get
$$
\hat G= \pmatrix{\hat G_{2''}^1 \cr \hat G_{2''}^2}
$$
and $\delta^{l}$.

\subsection{Representation of the one recursive step}

We can represent one recursive step as the following reduction of the matrix
$\tilde A$:
$$\tilde A=\pmatrix{A^1 \cr A^2 \cr}
 \rightarrow_1\pmatrix{\delta^{s} I_{s-k} & G_2^1 \cr
                           A_1^2       & A_2^2    }
 \rightarrow_2 \pmatrix{\delta^{s} I_{s-k} & G_2^1       \cr
                            0          & \hat A_2^2    }
\rightarrow_3
$$
$$
\rightarrow_3
\pmatrix{\delta^sI_{s-k} & G_{2'}^1 & G_{2''}^1    \cr
                           0 & \delta^{l} I_{l-s}& \hat G_{2''}^2 }
\rightarrow_4 \pmatrix{\delta^{l} I_{s-k} & 0    & \hat G_{2''}^1    \cr
                           0 & \delta^{l} I_{l-s}& \hat G_{2''}^2 }
=\pmatrix{\delta^l I_{l-k} & \hat G }
$$

\section{The Proof of the Main Identities}

\subsection{The first matrix identity}

The second step of the algorithm is based on the following matrix identity:
$$
\delta^k A^{s,l,(s+1)}_{s,c}=\delta ^{s}
A^{s,l,(k+1)}_{s,c}- A^{s,l,(k+1)}_{k,s} \cdot G^{k,s,(s)}_{s,c}.
$$
So we must prove the next identities
for the matrix elements
$$
\delta^k a_{ij}^{s+1}=\delta^{s} a_{ij}^{k+1}-\sum_{p=k+1}^{s}
a_{ip}^{k+1} \cdot \delta_{pj}^{s},
$$
$i=s+1, \ldots, l; \,\, j=s+1, \ldots, c$.

Let $\sigma_{ij}^k$ denote the minors that will stand in the place of the
minors $\delta^k$ after the replacement the row $i$ by the row $j$.
An expansion of the determinant $ a_{ij}^{k+1}$ according to the column $j$ is
the following
$$
 a_{ij}^{k+1}= \delta^k  a_{ij}- \sum_{r=1}^k \sigma_{ri}^k a_{rj}
$$
Therefore we can write the next matrix identity
$$
\pmatrix{
1&0&\cdots&0&0&\cdots&0&0\cr
0&1&\cdots&0&0&\cdots&0&0\cr
\vdots&\vdots&\ddots&\vdots&\vdots&\ddots&\vdots&\vdots\cr
0&0&\cdots&0&0&\cdots&1&0\cr
-\sigma_{1i}^k&-\sigma_{2i}^k&\cdots&-\sigma_{ki}^k&0&\cdots&0&\delta^k\cr
}\cdot {\bf A}_{ij}^{s+1}=
$$ $$
= \pmatrix{
a_{11}&a_{12}&\cdots&a_{1,s}&a_{1j}\cr
a_{21}&a_{22}&\cdots&a_{2,s}&a_{2j}\cr
\vdots&\vdots&\ddots&\vdots&\vdots\cr
a_{s,1}&a_{s,2}&\cdots&a_{s,s}&a_{s,j}\cr
a_{i1}^{k+1}&a_{i2}^{k+1}&\cdots&a_{i,s}^{k+1}&a_{ij}^{k+1}\cr
}$$
Note that $a_{ip}^{k+1}=0$ for $p \leq k$.
Finaly we decompose the determinant of the right matrix according
to the last row and write the determinant identity correspondingly to
this matrix identity.

\subsection{The second matrix identity}

The fourth step of the algorithm bases on the matrix identity
$$
\delta^{s} G^{k,s,(l)}_{l,c}=\delta ^{l} G^{k,s,(s)}_{l,c}
- G^{k,s,(s)}_{s,l} \cdot G^{s,l,(l)}_{l,c}.
$$

So we must prove the next identities
for the matrix elements:
$$
\delta^{s} \delta_{ij}^{l}=\delta^{l} \delta_{ij}^{s}-
\sum_{p=s+1}^{l} \delta_{ip}^{s} \cdot \delta_{pj}^{l},
$$
$i=k+1, \ldots, s;\,\,\, j=l+1, \ldots, c$.

Let $\gamma_{j,i}^{s}$ denote the algebraic adjunct of element $a_{j,i}$
in the matrix ${\bf A}_{s,s}^{s}$.
An expansion of the determinant $\delta_{ip}^{s}$ according to the column $i$
is the following
$$
\delta_{ip}^{s} = \sum_{q=1}^s \gamma_{qi}^s a_{qp}
$$

Therefore we can write the next matrix identity:
$$
\pmatrix{
1&0&\cdots&0&0&\cdots&0&0\cr
0&1&\cdots&0&0&\cdots&0&0\cr
\vdots&\vdots&\ddots&\vdots&\vdots&\ddots&\vdots&\vdots\cr
0&0&\cdots&0&0&\cdots&1&0\cr
\gamma_{1i}^{s}&\gamma_{2i}^{s}&\cdots&\gamma_{s,i}^{s}&0&\cdots&0&0\cr
} {\bf A}_{ij}^{l+1}=
$$ $$
= \pmatrix{
a_{11}&a_{12}&\cdots&a_{1,l}&a_{1j}\cr
a_{21}&a_{22}&\cdots&a_{2,l}&a_{2j}\cr
\vdots&\vdots&\ddots&\vdots&\vdots\cr
a_{l,1}&a_{l,2}&\cdots&a_{l,l}&a_{l,j}\cr
\delta_{i1}^{s}&\delta_{i2}^{s}&\cdots&\delta_{i,l}^{s}&
\delta_{ij}^{s}\cr
}$$
Note that $\delta_{ip}^{s}=0$ for $p \leq s$ and $\delta_{ii}^{s}=
\delta^{s}$. So to finish the proof
we must decompose the determinant of the right matrix according
to the last row and write the determinant identity correspondingly to
this matrix identity.

\section{Evaluation of Operations Number}

Let us have a method for matrix multiplications with the complexity $M(n)=
O(n^{2+ \beta})$, then for multiplication of two matrixes of order $l\times n$
and $n\times c$ we need $M(l \times n, n \times c)=O(lcn^{\beta})$ operations.
Let us denote by $S(n,m)$ the complexity of the recursive algorithm for the
matrix $A~\in~{\bf R}^{n\times m}$.

If in the first recursive step upper submatrix consists of the $s$ rows,
$1\leq s<n$, then
$$
S(n,m)=S(s,m)+M((n-s) \times s, s \times (m-s))+
$$
$$
+S(n-s,m-s)+M(s \times (n-s), (n-s) \times (m-n))+O(nm).
$$
For a matrix with $k$ rows we can choose the arbitrary $s: 1 \leq s \leq k-1$.

If the process of partition is dichotomous, and the number of rows in the
upper and lower submatrixes is the same in every step, then
$S(2n,m)$ satisfies the recursive inequality:
$$ S(2n, m) = S(n,m)+M(n \times n, n \times (m-n))+ S(n,m-n)+
$$ $$
+M(n \times n, n \times (m-2n))+ O(nm) \leq 2S(n,m)+2O(m n^{\beta +1}).
$$
So we have
$$
S(2n,m) \leq nS(2,m)+ \sum ^{(\log_2n)-1}_{i=0} O(({n \over 2^i})^
{\beta +1} m) 2^{i+1}=
$$
$$
=nS(2,m)+{2 \over 1-2^{-\beta}}O((n^ \beta -1) n m)
$$
And finally
$$
S(2n,m) \leq O(m n^{\beta +1}).
$$
On the other hand
$$
S(2n,m) > M(n \times n, n \times (m-n))=O(m n^{\beta +1}).
$$
Therefore
$$
S(2n,m)=O(m n^{\beta +1}).
$$
So the complexity of this algorithm is the same that the complexity of the
matrix multiplication. In particular for $m=n+1$ we have
$$
S(n,n+1)=O(n^{2+ \beta})
$$

It means that the solution of the system of linear equations
needs (accurate to the constant multiplier) the same number
of operations that the multiplication of two matrixes needs.

We can get the exact number of operations, that are necessary for the
solution of the system of linear equations of order $n \times m$,
in the case when on every step upper submatrix is no less then lower
submatrix and the number of rows in upper submatrix is some
power of $2$.

Let
$F(s,\mu-s,\nu)= M((\nu-s) \times s, s \times (\mu-s))+
M(s \times (\nu-s), (\nu-s) \times (\mu-\nu))$, then we obtain $ S(n,m)$:
$$
\sum_{k=1}^{\lfloor \log_2 n \rfloor}
(F(2^k,n-2^k \lfloor {n \over 2^k} \rfloor,
m - 2^k(\lfloor {n \over 2^k}\rfloor -1)) +
\sum_{i=1}^{\lfloor n/2^k \rfloor}
F(2^{k-1},2^{k-1}, m -(i-1) 2^k))
$$
Let $n=2^p$. If we use simple matrix multiplications with
complexity $n^3$ than we obtain
$$A_{nm}=(6n^2m-4n^3-6nm+3n^2+n)/6,$$
$$M_{nm}=(6n^2m-4n^3+(6nm-3n^2)\log_2 n-6nm+4n)/6,$$
$$D_{nm}=((6nm-3n^2)\log_2 n - 6nm-n^2+ 6m+3n-2)/6.$$
Here we denote by $A_{nm}, M_{nm}, D_{nm}$  the numbers of
additions/subtractions, multiplications and divisions, and
take into account that $(6nm - 2n^2 - 6m +2)/6$ divisions
in the second step are divisions by $\delta^0=1$, so they do
not exist in $D_{nm}$.

For m=n+1 we obtain
$$A_{n,n+1}=(2n^3+3n^2-5n)/6,$$
$$M_{n,n+1}=(2n^3+(3n^2+6n)\log_2 n -2n)/6,$$
$$D_{n,n+1}=(3n^2\log_2 n - 7n^2+ 6n\log_2 n +3n+4)/6.$$
The general quantity of multiplication and division operations
is about $n^3/3$.

We can compare these results with one-pass algorithm,
that was the best of all known algorithms (see \cite{mal-1991}):
$A^O_{n,n+1}=(2n^3+3n^2-5n)/6$,
$M^O_{n,n+1}=(n^3+2n^2-n-2)/2$,
$D^O_{n,n+1}=(n^3 - 7n+ 6)/6$,
the general quantity of multiplication and division operations
is about $2n^3/3$.

If we use Strassen's matrix multiplications with
complexity $n^{\log_2 7}$ than  we can obtain for $n=2^p$
the general quantity of multiplication and division operations
$$ MD^S_{n,m}=
n^2(\log_2 n -5/3) +7/15n^{\log_2 7}+(m-n)(n^2 + 2n\log_2 n  -n) + 6/5n-$$
$$-\sum_{i=1}^{\log_2 n -1} {n\over 2^i}(8^i-7^i)
\{ \lfloor{m-n \over 2^i}\rfloor - ({n \over 2^i}-2)
\lfloor {m-n-2^{i+1}\lfloor(m-n)/ 2^{i+1}\rfloor  \over 2^i} \rfloor\}.$$
For $m=n+1$, $n=2^p$ we get
$$ MD^S_{n,n+1}={7\over 15}n^{\log_2 7}+n^2(\log_2 n-{2\over 3})
+n(2\log_2 n +{1\over 5}). $$

\section{Example}

Let us consider next system over the integer numbers
$$
\pmatrix{3&1&1&-1\cr
         1&2&0& 1\cr
         0&1&2& 0\cr
         1&0&0& 2\cr} \cdot
\pmatrix{ x_1\cr x_2\cr x_3\cr x_4\cr } =
\pmatrix{ 4\cr 4\cr -2\cr -1\cr }
$$

\subsection{Reduction of the matrix $A^1=A_{05}^{02(1)}$ to the diagonal form}

We make the next reduction:
$$A_{05}^{02(1)} \rightarrow (\delta^2 I_2,~G_{25}^{02(2)} )$$

\subsubsection{}
$$A_{05}^{01(1)} \rightarrow (\delta^1 I_1,~G_{15}^{01(1)} ) = (3;1,1,-1,4)$$

\subsubsection{}
$$
\delta^0 A_{15}^{12(2)}=\delta^1 A_{15}^{12(1)}-A_{01}^{12(1)} G_{15}^{01(1)}=
$$
$$ = 3(2,0,1,4)-(1)(1,1,-1,4) = (5,-1,4,8),~~~\delta^0 \equiv 1. $$

\subsubsection{}
$$A_{15}^{12(2)} \rightarrow (\delta^2 I_1,~G_{25}^{12(2)} ) = (5;-1,4,8)$$

\subsubsection{}
$$\delta^1 G_{25}^{01(2)}= \delta^2 G_{25}^{01(1)}-G_{12}^{01(1)}
G_{25}^{12(2)}=$$
$$= 5(1,-1,4) - (1)(-1,4,8) = (6,-9,12)$$
$$ G_{25}^{01(2)}=(2,-3,4)$$

Finally we obtain
$$(\delta^2 I_2,~G_{25}^{02(2)})=
\pmatrix{5&0;&2&-3&4\cr
         0&5;&-1&4&8\cr}$$

\subsection{Computation of the matrix $\hat A_2^2=A_{25}^{24(3)}$}

$$\delta^0 A_{25}^{24(3)}=\delta^2 A_{25}^{24(1)}-A_{02}^{24(1)}
G_{25}^{02(2)}=$$
$$=5\cdot \pmatrix{2&0&-2\cr 0&2&-1\cr}-\pmatrix{0&1\cr 1&0\cr}
\pmatrix{2&-3&4\cr -1&4&8\cr}=
$$ $$
=\pmatrix{11&-4&-18\cr -2&13&-9\cr}
$$
$$
\delta ^0 \equiv 1; \;\; A_{25}^{24(3)}=\pmatrix{11&-4&-18\cr -2&13&-9\cr}
$$

\subsection{
Reduction of the matrix $A^2_2=A_{25}^{24(3)}$ to the diagonal form}

We make the next reduction:
$$
 A_{25}^{24(3)} \rightarrow (\delta^4 I_2~G_{45}^{24(4)} )
$$

\subsubsection{}
$$A_{25}^{23(3)} \rightarrow (\delta^3 I_1,~G_{35}^{23(3)} ) = (11;-4,-18)$$

\subsubsection{}
$$
\delta^2 A_{35}^{34(4)}=\delta^3 A_{35}^{34(4)}-A_{23}^{34(3)} G_{35}^{23(3)}=
$$
$$ = 11(13,-9)-(-2)(-4,-18) = (135,-135) $$
$$A_{35}^{34(4)}=(27,-27)$$

\subsubsection{}
$$A_{35}^{34(4)} \rightarrow (\delta^4 I_1,~G_{45}^{34(4)} ) = (27,-27)$$

\subsubsection{}
$$\delta^3 G_{45}^{23(4)}= \delta^4 G_{45}^{23(3)}-G_{34}^{23(3)}
G_{45}^{34(4)}=$$
$$=27(-18) - (-4)(-27) = -594,~~  G_{45}^{23(4)}= (-54)$$

Finally, in step (3) we obtain
$$(\delta^4 I_2,~G_{45}^{24(4)})=
\pmatrix{27&0;&-54\cr
         0&27;&-27\cr}$$

\subsection{
Computation of the matrix $\hat G_{2''}^1=G_{45}^{02(4)}$}

$$\delta^2 G_{45}^{02(4)}= \delta^4 G_{45}^{02(2)}-G_{24}^{02(2)}
G_{45}^{24(4)}=$$
$$=27\pmatrix{4 \cr 8\cr} - \pmatrix{2&-3\cr -1&4 \cr}
\pmatrix{-54 \cr -27 \cr}=\pmatrix{135 \cr 270 \cr}
$$
$$
G_{45}^{02(4)}=\pmatrix {27 \cr 54 \cr}
$$
The solution of the system is the following:
$$
\delta^4=27;~~~G_{45}^{04(4)}=\pmatrix {27 \cr 54 \cr -54 \cr -27 \cr}
$$

\subsection{Representation of the first recursive step}

We can represent the first recursive step as the following

$$
A \rightarrow_1
\pmatrix{5&0&2&-3&4\cr
         0&5&-1&4&8\cr
         0&1&2& 0&-2\cr
         1&0&0& 2&-1\cr}
\rightarrow_2
\pmatrix{5&0& 2&-3& 4\cr
         0&5&-1& 4& 8\cr
         0&0&11&-4&-18\cr
         0&0&-2&13&-9\cr}
\rightarrow_3
$$ $$
\rightarrow_3
\pmatrix{5&0& 2&-3& 4\cr
         0&5&-1& 4& 8\cr
         0&0&27&0&-54\cr
         0&0&0&27&-27\cr}
\rightarrow_4
\pmatrix{
         27&0& 0& 0& 27\cr
         0&27& 0& 0& 54\cr
         0& 0&27& 0&-54\cr
         0& 0& 0&27&-27\cr}$$

\section{Conclusion}

The described algorithm for the solution of the systems of linear equations
over the integral domain includes the known one-pass method and the method
of forward and back-up procedures \cite{mal-1991}.
If in every recursive step the partition of the matrix is such that
the upper submatrix consists only of one row then it is the method  of
forward and back-up procedures. If the lower submatrix consists only of
one row in every step then it is the one-pass method.

If the process of partition is dichotomous and the numbers of rows in the
upper and lower submatrixes are equal in every step, then the
complexity of the solution has
the same order $O(n^{2+ \beta})$ as the complexity of matrix multiplication.

The computation of the matrix determinant and the computation of
the adjugate matrix have the same complexity.

This method may be used in any commutative ring if the corner minors
$\delta^k,~k=1,2,\ldots,n$, do not equal zero and are not zero divisors.

\bibliographystyle{plain}

\begin{thebibliography}{STR-1999}

\bibitem{str-1969}
  V.\ Strassen.
  Gaussian Elimination is not optimal.
  {\em Numerische Mathematik}, 1969, {\bf 13}, 354--356.

\bibitem{cw-1987}
  D.\ Coppersmith,\ S.\ Winograd.
  in {\em Proc. 19th Annu ACM Symp. on Theory of Comput.}, 1987, 1--6.

\bibitem{dod-1866}
  C.L.\ Dodgson.
  Condensation of  determinants,  being  a  new  and brief method
  for computing their  arithmetic  values.
  {\em Proc.  Royal Soc. Lond.}, 1866, {\bf A.15}, 150--155.

\bibitem{bar-1968}
  E.N.\ Bareiss.
  Sylvester's identity and multistep integer-preserving
  Gaussian elimination.
  {\em Math. Comput.}, 1968, {\bf 22}, 565--578.

\bibitem{mal-1983}
  G.I.\ Malaschonok.
  Solution of a system of linear equations in  an integral domain.
  {\em USSR  Journal  of  Computational  Mathematics  and Mathematical
  Physics}, 1983, {\bf 23}, 1497--1500.

\bibitem{mal-1987}
  G.I.\ Malaschonok.
  On the solution of a linear equation system over commutative ring.
  {\em Math. Notes of the Acad. Sci. USSR}, 1987, {\bf 42}, {\bf N4}, 543--548.

\bibitem{mal-1989}
  G.I.\ Malaschonok.
  A  new  solution  method  for  linear  equation systems over  the
  commutative  ring. In {\em  Int. Algebraic Conf., Theses on the ring theory,
  algebras and modules.} Novosibirsk, 1989, 82--83.

\bibitem{mal-1991}
  G.I.\ Malaschonok.
  Algorithms for the solution of systems of linear equations in commutative
  rings. In {\em Effective Methods in  Algebraic Geometry}, Edited by
  T. Mora  and C. Traverso,  Progress  in Mathematics 94,
  Birkhauser, Boston-Basel-Berlin, 1991, 289--298.

\end{thebibliography}

\end{document}